\documentclass[lettersize,journal]{IEEEtran}
\usepackage{amsmath,amsfonts}
\usepackage{algorithmic}
\usepackage{algorithm}
\usepackage{array}
\usepackage{caption}
\usepackage[caption=false,font=normalsize,labelfont=sf,textfont=sf]{subfig}
\usepackage{textcomp}
\usepackage{stfloats}
\usepackage{url}
\usepackage{verbatim}
\usepackage{graphicx}
\usepackage{cite}
\usepackage[nolist,nohyperlinks]{acronym}
\usepackage{makecell}
\usepackage{color}

\newcommand{\upd}[1]{\textcolor{black}{{#1}}}

\begin{document}
\bstctlcite{IEEEexample:BSTcontrol}
\begin{acronym}
    \acro{ISAC}{integrated communication and sensing}
    \acro{BS}{base station}
    \acro{UE}{user equippment}
    \acro{JSAC}{joint sensing and communication}
    \acro{LOS}{line-of-sight}
    \acro{NLOS}{non-line-of-sight}
    \acro{TRP}{transmission and reception point}
    \acro{C2S}{communication-to-sensing}
    \acro{S2C}{sensing-to-communication}
    \acro{MIMO}{multiple-input multiple-response}
    \acro{D-MIMO}{Distributed \ac{MIMO}}
    \acro{IAB}{integrated access and backhaul}
    \acro{IRS}{intelligent reflecting surface}
    \acro{NCR}{network controller repeater}
    \acro{DL}{downlink}
    \acro{UP}{uplink}
    \acro{Tx}{transmitting}
    \acro{Rx}{receiving}
    \acro{DoA}{direction of arrival}
    \acro{SNR}{signal-to-noise ratio}
    \acro{SINR}{signal-to-interference-plus-noise ratio }
    \acro{SI}{self-interference}
    \acro{gNB}{gNodeB}
\end{acronym}
\title{Mobility Management in Integrated Sensing and Communications Networks}

\author{Yuri S. Ribeiro, Behrooz Makki, André L. F. de Almeida, Fazal-E-Asim, Gabor Fodor
}

\markboth{Journal of \LaTeX\ Class Files,~Vol.~14, No.~8, August~2021}%
{Shell \MakeLowercase{\textit{et al.}}: A Sample Article Using IEEEtran.cls for IEEE Journals}


\maketitle

\begin{abstract}
The performance of the integrated sensing and communication (ISAC) networks is considerably affected by the mobility of the transceiver nodes, {user equipment devices} (UEs) and the {passive objects that are sensed}.
For instance, the sensing efficiency is considerably affected by {the presence or absence} of a line-of-sight connection between the sensing transceivers and the object; a condition that may change quickly due to mobility.
Moreover,  the {mobility of the UEs and objects} may result in {dynamically} varying communication-to-sensing and sensing-to-communication {interference}, deteriorating the network performance. In such cases, there may be a need to handover the sensing process to neighbor nodes. In this article, we develop the concept of mobility management 
in ISAC networks. Here, depending on the {mobility of} objects and/or the transceiver nodes, the data traffic, the sensing or communication coverage area of the transceivers, and the network interference, the transmission and/or the reception of the sensing signals may be {handed over} to neighbor nodes. 
Also, the ISAC configuration and modality -- {that is, using monostatic or bistatic sensing} -- are updated accordingly, such that the sensed objects can be continuously sensed with low overhead. We show that mobility management {reduces} the sensing interruption and boosts the communication and sensing efficiency of ISAC networks. 
\end{abstract}


\section{Introduction}
\IEEEPARstart{F}{\lowercase{or}} decades, wireless telecommunications and radar {technologies} have co-existed, and most of the effort so far has been on interference management so that these two technologies can co-exist with limited interference to each other. Currently and in the near future, millimeter wave (mmwave) communication bands are merging with the radar bands for high-resolution sensing, allowing for efficient spectrum usage by combining both functionalities. Moreover, because of the high reliance on beamforming for communication in high frequencies, the need for spatial awareness and accurate positioning of objects connected or not to the network has increased significantly in 5G and beyond. These open up opportunities for the so-called \ac{ISAC}.


With ISAC, the same spectrum and/or wireless network infrastructure are used for both sensing and communication functionalities. Here, the term sensing refers to \emph{radar-like} functionalities, i.e., the capability to detect the presence and to track the movement and other characteristics of connected or unconnected objects in the wireless network coverage area. Compared to the deployment of a separate network for sensing functionality, the main advantages of ISAC are:
\begin{itemize}
    \item The sensing functionality is introduced on a large scale at a relatively low incremental cost by reusing the infrastructure that is already deployed for communication purposes.
    \item Having communication and sensing under the network control simplifies the resource allocation, interference management, and regulation aspects.
    \item Introducing the sensing improves the network knowledge about the area, which, in turn, helps in blockage avoidance, backhaul requirement reduction, timely channel state information acquisition, accurate beamforming, etc.
\end{itemize}

A review of ISAC networks can be found in, e.g., \cite{Guo2024integrated,Zhang2021Overview,Zhang2022Enabling,Liu2022Integrated}. Moreover, initial test-bed evaluations of ISAC can be found in, e.g., \cite{Guo2024integrated}, \cite{Zhang2021Design,Colpaert2023Massive,Ji2023Networking}. Regarding standardization, 3GPP has recently started a study item on ISAC in Release 19 focusing on channel modeling \cite{3gpprel19}, and ISAC is considered as one of the key 6G candidates. Particularly, 6G is considered as a multi-functional network with all communication, localization, and sensing functionalities in the same spectrum/hardware.

Sensing methods can be divided into two categories: mono-static and multi-static. With mono-static sensing, the same node transmits and receives the sensing signal. With multi-static sensing, on the other hand, the transmission and the reception are handled by different collaborating nodes. 
Mono-static sensing requires a full duplex and high \ac{SI} cancellation capabilities. Multi-static sensing, on the other hand, requires tight coordination and synchronization between the cooperative network nodes. 

In one view, ISAC is an enabling technology
for perceptive cellular networks, and it is natural that perception needs to work well
in the presence of mobility, where different nodes in ISAC network, including the object itself, the \acp{UE} and the communication/sensing transceivers may be mobile \cite{ZhangPMN2021}. This fundamental difference motivates to rethink mobility management techniques in ISAC networks. 

With ISAC, one may consider two scenarios where either all sensing transceivers are stationary, e.g., \acp{gNB}, or the case where one or more of the sensing transceivers is mobile, e.g., UE. The cases with one or more of the sensing transceivers being a UE is of interest as it has less dynamic time division duplexing or full-duplex requirements, simplifying the interference management. Moreover, compared to a gNB, the UE is normally closer and at a similar height to the objects, which increases the \ac{LOS} probability. 
On the other hand, involving the UEs in the sensing process has disadvantages, including lower transmit power and beamforming/signal processing capabilities at the UE, compared to the gNBs. Moreover, compared to the cases with stationary sensing gNBs, the UEs' sensing coverage area is small, the network performance is less predictable, and there may be a need for frequent handovers.  Moreover, due to the UE’s mobility, there may be a need for joint UE localization and object sensing, which affects the sensing performance. Finally, different from the cases with UEs, stationary gNBs may have access to high-rate, e.g., fiber, backhaul links, which facilitate the coordination and synchronization significantly.  

Different from communication networks, which can operate well in cases with \ac{NLOS} connections between the gNB and the UE, the efficiency of the sensing systems depends a lot on the presence of LOS links between both the \ac{Tx} sensing node and the object as well as between the object and the \ac{Rx} sensing node (although there are limited works on sensing in \ac{NLOS} conditions, e.g., \cite{Gonzalez2024Integrated}). 
However, different measurements show a low probability of LOS, especially in dense areas where, for instance, considering a London scenario, the LOS probability may be down to around $27\%$ and $20\%$ in the cases with mono-static and multi-static cases, respectively \cite{MollenMultistatic}. Here, the LOS may be lost quickly due to the transceivers and/or the objects' mobility. Moreover, due to the nodes'/objects' mobility, ISAC may result in dynamic \ac{C2S} and/or \ac{S2C} interferences on top of existing communication-based interferences. In all such cases, mobility management may be needed. 

Historically, there are two types of mobility in cellular networks for idle and connected mode UEs. In practice, using the idle mode signals for sensing may be cumbersome and require extensive standardization efforts. For this reason, in this paper, we concentrate on connected mode mobility where the sensing process may be \emph{handovered} to the neighbor nodes. In the proposed scheme, handover events are defined where an occured event triggers handovering the sensing process to neighbor nodes, enabling mobility management in ISAC networks.

Sensing handover is of particular interest because one of the key points of advanced 5G and 6G is network densification \cite{Madapatha2020Integrated} in which the area will be densified with multiple nodes of different capabilities, which will give a chance for proper sensing handover. Particularly, distributed multi-input multi-output networks, with multiple cooperative multi-antenna nodes in an area, are foreseen to play an important role in 6G. 

With this background, this paper studies the potentials and challenges of mobility management and, in particular, sensing handover in ISAC networks. Once an object cannot be properly sensed (or the network predicts that it cannot be sensed soon) due to, e.g., mobility, high network interference, etc., the network may request the other (target) network nodes to hand over the object's sensing to them. 
Here, not only the transmission and/or the reception of the sensing signals may be handovered to neighbor nodes, but also the modality of the sensing system may be changed where, for example, we switch between different 
downlink (DL)/uplink (UL) mono-static and/or multi-static 
sensing methods. 

As we explain in the following, while sensing mobility management is built on the same fundamentals as communication mobility management, there are conceptual differences between them that drastically affect the design of sensing mobility management.
With sensing, the triggering mechanism and the signaling to support the handover differ from the communication since the object might not be connected to the network, and the information to be shared with the cooperative nodes may be different from those with communication-based handover.
Finally, the objectives for sensing handover may differ fundamentally from those in communication-based handover. As a result, the adaptation protocols at the Tx and Rx nodes need to be inherently different from the ones adapted for typical communication-based mobility management. 

We investigate cases with soft sensing handover where the object may be sensed by different entities for an overlapping time period. The case study results are also presented as proof of concept.
As illustrated, with proper configurations, sensing handover enables continuous sensing and tracking of the objects in ISAC networks. This, in turn, improves the network's spatial awareness, which results in proper communication as well. Moreover, as explained in the following, sensing handover avoids strong C2S and S2C interferences, resulting in robust communication and sensing. 




\section{Sensing Handover in ISAC Networks}

In wireless communications, handover refers to the process of transferring an ongoing communication from one cell to another without interruption. Handovers are crucial for maintaining the quality and continuity of communication as UEs move through different network areas. There are different types of handovers, including handovers within the same frequency band, between different frequency bands, and between different radio access technologies, such as Global System for Mobile Communications to Universal Mobile Telecommunication System.

In 5G new radio (NR), communication handover typically involves several signaling steps between the UE, the source gNB (gNB from which the UE is moving away), and the target gNB (gNB to which the UE is moving).
The physical layer and radio resource control specifications for communication handover can be found on \cite{TS214} and \cite{TS331}.

One of the benefits of ISAC networks is that wireless communication techniques, which have been already developed and
optimized over the decades, can be extended for sensing applications. With this motivation, in this paper, we introduce the sensing handover concept in ISAC networks as follows.

\begin{figure*}[!t]
    \centering
    \includegraphics[width=18cm,height=10cm,keepaspectratio]{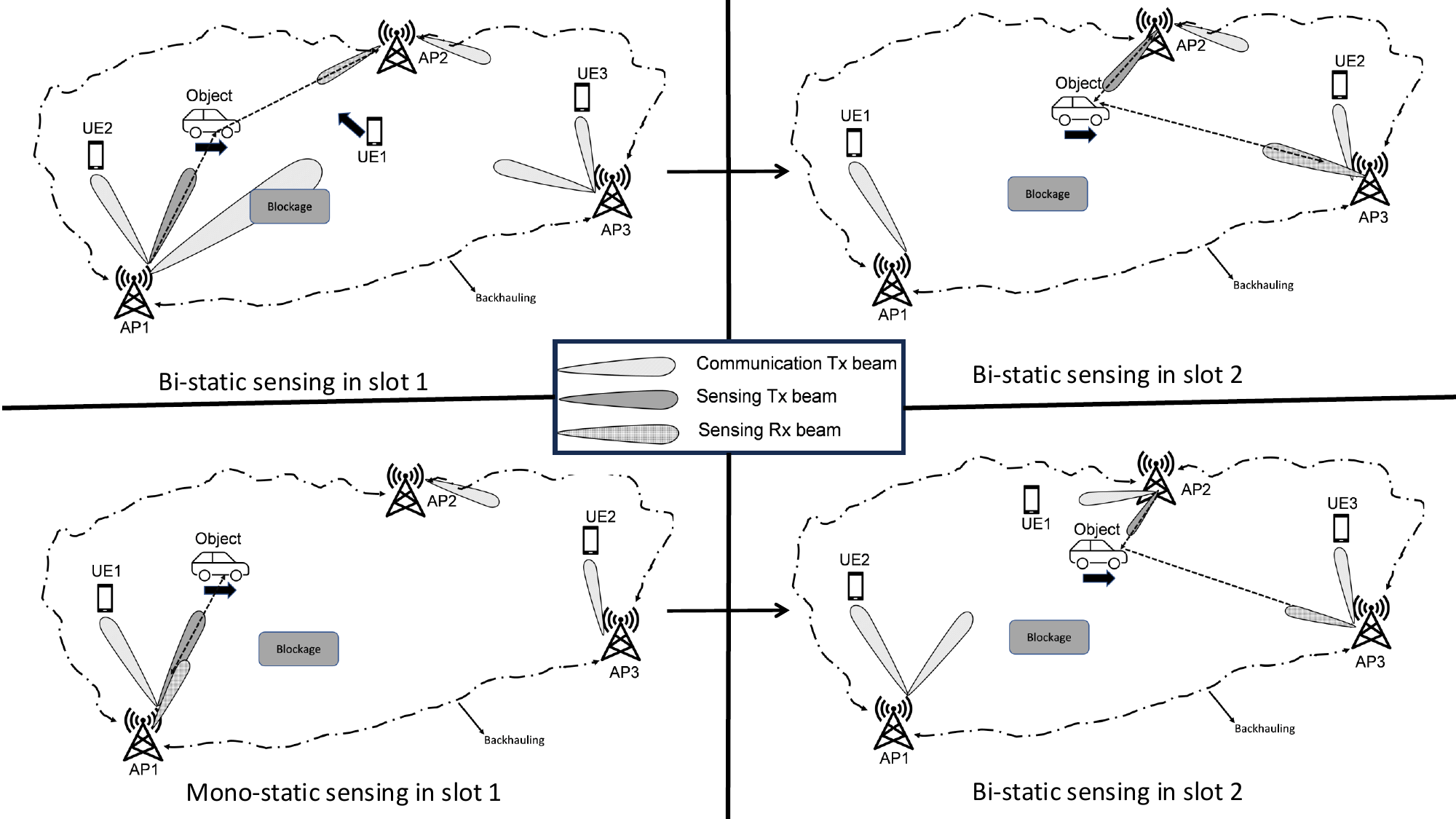}
    \caption{Sensing handover illustrative scenarios.}
    \label{fig:illustration}
\end{figure*}

Consider the ISAC networks of Fig. 1, with mono-static and bi-static sensing. In Slot 1, an access point (AP), e.g., AP1, performs the ISAC process either in cooperation with AP2 (bi-static sensing) or alone (mono-static sensing). In general, the AP may refer to a gNB or UE, depending on the considered ISAC network (although UE-based mono-static sensing is not of interest in practice). As explained in the following, for different reasons, there may be a need to handover the transmission and/or the reception of sensing functionality to other network nodes. Here, the sensing handover is based on the following steps: 
\begin{itemize}
    \item Determining the need for sensing handover based on predefined criteria, 
\item Performing the sensing handover based on the obtained determination. 
\end{itemize}
Note that, in conjunction with the handover decision, the modality of the sensing, i.e., the mono-static/multi-static type, DL/UL-based sensing method, or even the transmission/reception functionality of the nodes, may also be changed. 
\subsection{Determining the Need for Sensing Handover}

In a typical wireless network, handover is event-based, i.e., a handover process is triggered dynamically when an event occurs, e.g., the received signal power at a target node exceeds the received power of the source node by a threshold. However, long-term conditions may also result in sensing handover, along with dynamic events. Moreover, not only the received signal power but also various dynamic events may result in sensing handover. Long-term conditions triggering the sensing handover include: 
\begin{itemize}
    \item Sensing coverage area: Each node may have a limited sensing coverage area when operating as a sensing signal transmitter or receiver, and the sensing coverage area 
    is mainly determined by the area guaranteeing LOS connection to the node.
    As a result, if the object is estimated or predicted to move out of the sensing coverage of a Tx or Rx node,  a handover request may be initiated. 
    \item Beam restrictions (in both cases where a node is in the Tx and/or the Rx mode of sensing): With ISAC, not all combinations of the Tx and Rx beams are necessarily available for simultaneous sensing and communication. The beam restrictions may be determined during the network planning (in the cases with stationary sensing transceivers) to avoid strong interference, mainly in cases where the APs are operating in full-duplex mode.
    Considering the beam restrictions, if the object moves towards the areas covered by the restricted beams, the sensing needs to be handovered to other nodes.
\end{itemize}
Along with long-term conditions, which are determined during network planning, there may be dynamic events triggering sensing handover:  
\begin{itemize}
    \item Resource allocation trade-off: With ISAC, the limited time, frequency, and/or energy resources must be efficiently allocated between the sensing and communication functionalities. 
    Then, due to the dynamic traffic/mobility of the UEs, it is probable that a node may be unable to continue allocating resources for sensing, triggering a sensing handover. 
    \item Dynamic blockage and deep fade: Assume a highway scenario where the gNBs track the vehicles. If a vehicle is blocked (or is predicted to be blocked), other cooperative nodes need to continue the sensing process. Such a problem becomes even more important in cases where the UEs are the sensing transceiver due to high and unpredictable UE mobility.
    \item Interference management: Consider the cases where a communicating UE and an object are close to each other. Here, high C2S and/or S2C interference may affect the communication and/or sensing functionalities, and, therefore, the network may decide to hand over the sensing to other nodes. Moreover, the UEs may demand high throughput and/or low latency communication, and in such cases, the network may prefer to keep the communication link interference-free.
    \item Sensing resolution: Depending on the object's positions, two or more objects may have similar angles of arrival, and, thereby, the receiver may not be able to estimate them accurately, while it can understand the presence of more than one object in the area. Here, one solution is to hand over part of the sensing process to another node or even switch the Tx/Rx mode of operation of the sensing nodes.
\end{itemize}
In this way, the need for sensing handover may be understood either at the current sensing transceivers (e.g., in the cases with high observed interference, received power drop, beam restriction, etc.) or new candidate transceivers for sensing handover (e.g., in the cases with finding better candidate nodes, resource allocation trade-off, etc.), depending on the triggering event. This implies that both the current and candidate set of transceivers need to be aware of the spectrum resources in which the object is currently sensed and perform measurement. This is, however, possible given that, depending on the network deployment, there may be few transceivers in an area with the possibility of sensing handover.

Different from handover in wireless communication, in which a UE is handovered from a source gNB to a target gNB, in multi-static sensing handover, one or both the sensing transmitter and receiver may be switched during the handover process.  
Furthermore, to determine whether the sensing handover should be at the Tx and/or Rx network node, the following aspects must be taken into account 1) the sensing and/or communication load at the network nodes and their priorities, 2) DL/UL configurations, 3) (self-)interference measurements at different nodes, and 4) knowledge about the area. 

\subsection{Performing the Sensing Handover}
Once the need for sensing handover is understood, the network initiates the sensing handover process. For instance, considering Fig. 1, AP1 may request one or more of the target network nodes, e.g., AP3, for sensing handover either as a sensing transmitter or receiver. Here, either a handover request may be broadcasted to multiple nodes, or dedicated handover requests may be sent to specific APs (i.e., gNBs or UEs), depending on, e.g., the object moving trajectory, the AP's level of cooperation/synchronization, etc. Accordingly, AP1 may receive feedback from one or more of the target network nodes indicating whether the sensing handover request is received and whether the target node accepts the request.

In one alternative, either before or after the sensing handover request, the network may provide the target network node(s) with assisting information where the assisting information may include information about, e.g., the moving trajectory, the speed, the position or the characteristics (size, etc.) of the object, information about the area, the communication/sensing coverage area of the other cooperative nodes, the beam restrictions of cooperative nodes, the available resource (e.g., frequency bands, time resources, maximum Tx power) of the cooperative nodes, etc. Moreover, the assisting information may indicate whether the previous Tx/Rx nodes are still available/useful to be involved in the sensing process of the object or not. For instance, in cases where AP2 is blocked, AP1 may still be available to sense the object. Also, in the cases with mono-static sensing and when an AP is equipped with physically separated panels/sub-arrays, while the Rx panels/sub-arrays may be blocked, the Tx panels/sub-arrays may still have a good view of the object, or vice versa. 

The assisting information simplifies the sensing process at the target node. For instance, knowing the rough location of the object allows the target node to skip beam sweeping over multiple wide beams and initiate the sensing process with narrow beams pointing towards the object. As a result, not only is the sensing latency reduced, but also the additional S2C interferences are avoided.

Once the assisting information is received, the target sensing transceiver determines whether it accepts or not the handover. Accordingly, if it accepts the handover, appropriate configuration changes are applied at the new sensing transceivers, and a handover configuration message is sent to the network. Note that, in one alternative, all decisions and re-configurations are performed at a central unit. In a distributed setup, on the other hand, each node may determine its appropriate configurations, e.g., based on the received assisting information, and inform the other cooperative nodes accordingly.

Soft handover is a well-defined concept in wireless communications, where the UE is connected to two gNBs for a short period before it is fully handovered to the target gNB. This guarantees uninterrupted communication for the UE. The same approach is applicable in sensing handover. With soft handover, upon initiating the sensing handover, e.g., the previous sensing receiver (source node) may still continue the sensing process with the same or different configurations for some time resources and inform the target network node(s) about the sensing results. Then, the information received from the source node can be combined with the sensing results obtained at the target node, improving the sensing accuracy. Here, the number of time resources to continue the sensing process at the source node may be pre-defined. Alternatively, soft handover may continue until the sensing performance at the source node drops below a predefined threshold or until the source node, e.g., AP2 in Fig. 1, receives an indication from the target network node(s), e.g., AP3 in Fig. 1, to terminate the sensing process.


\section{Case study}
\begin{figure}[t]
    \centering
    \includegraphics[width=\linewidth]{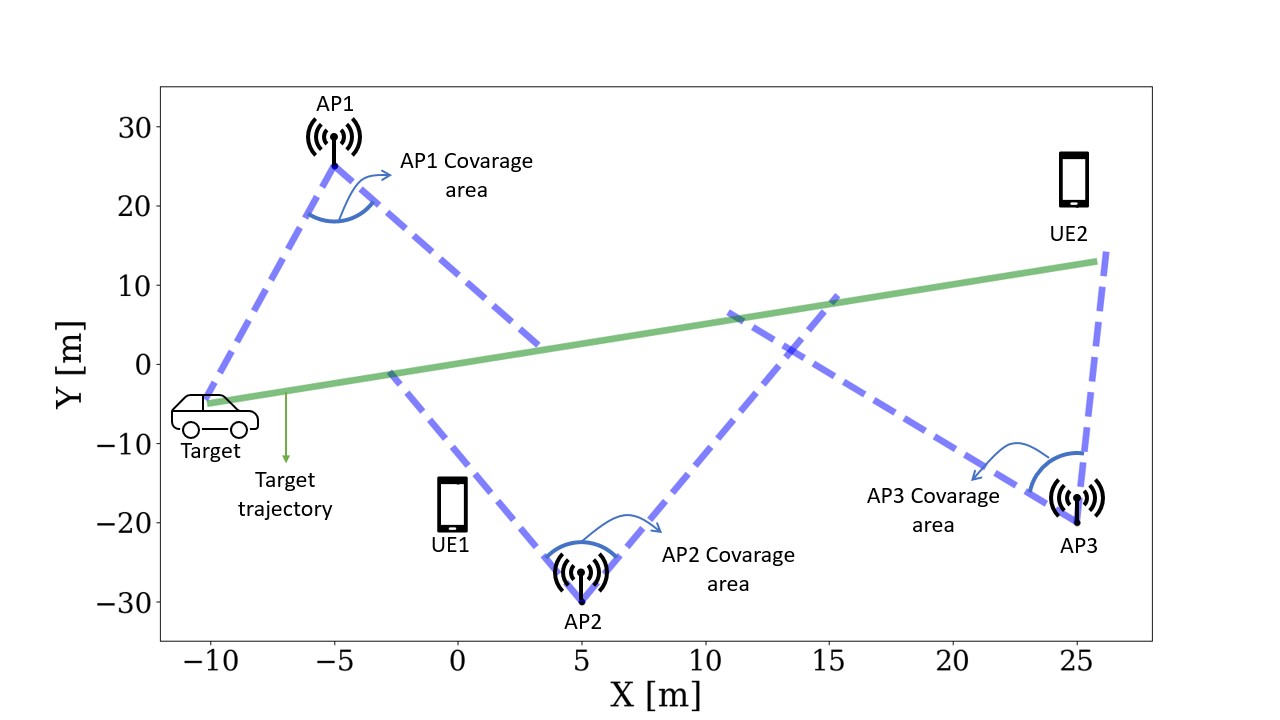}
    \caption{Case-study simulation setup}
    \label{fig:system model}
\end{figure}
To show the benefits of the mobility management, we consider a case study in an 
ISAC system where $3$ APs
collaborate to track one object moving from left to right as illustrated in Fig. \ref{fig:system model}. 
It is assumed that the APs can share information via, e.g., fiber backhaul or wireless control links. 
For the following results, $120$ snapshots are taken during the object's trajectory, and the measurements are realized at each snapshot. The simulation parameters are given in Table \ref{tab:parameters}. Also, the simulation codes can be found in \cite{YuriGit2024} for readers interested in reproducing the results.

\begin{figure*}[t]
    \centering
    \includegraphics[width=\linewidth,height = 10cm]{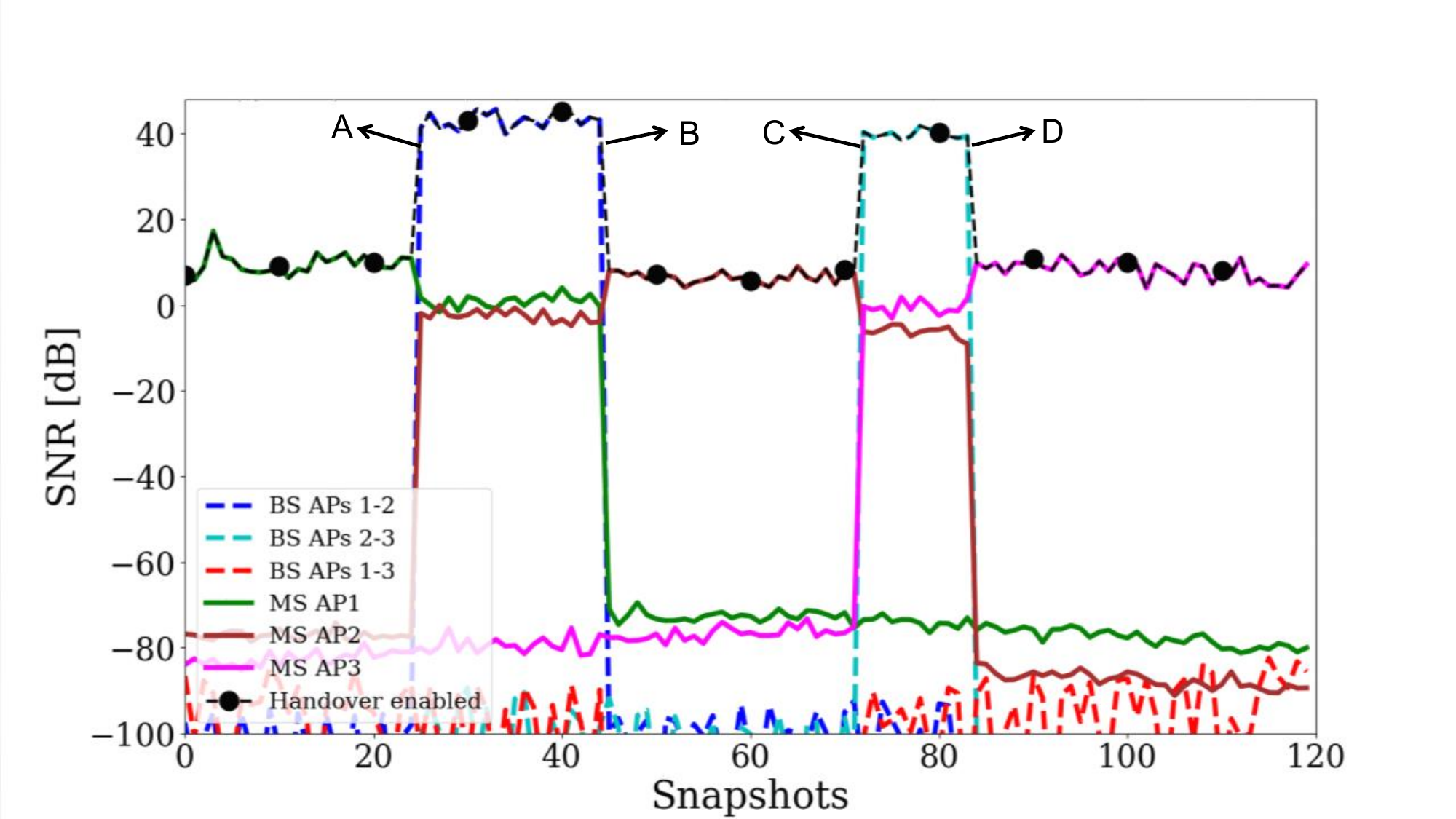}
    \caption{Sensing SNR as the object moves in an interference-free scenario. The handover SNR curve indicates the sensing configuration, and A, B, C, and D indicate different handovers. A: Mono-static (MS) with AP1 handovered to Bi-static (BS) with APs 1 and 2. B: BS with APs 1 and 2 handovered to MS with AP2. C: MS with AP2 handovered to BS with APs 2 and 3. D: Bi-static with APs 2 and 3 handovered to mono-static with AP3.}
    \label{fig:sensing scenario}
\end{figure*}

\begin{figure}[t]
    \centering
    \includegraphics[width=\linewidth]{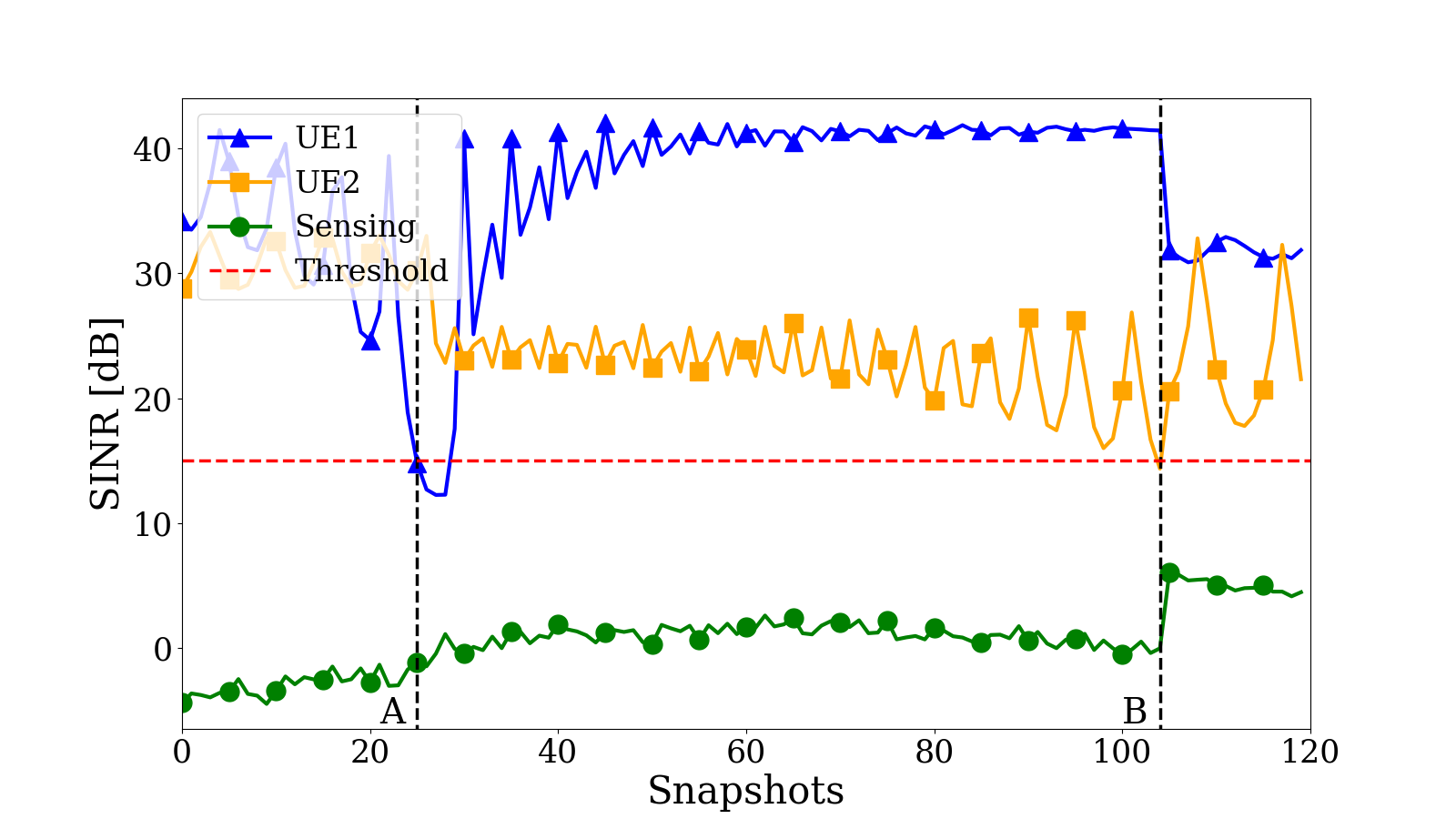}
    \caption{\upd{Communication and sensing SINRs as the object moves in a blockage-free scenario. UE1 and UE2 indicate the observed communication SINR to the network. The handover points are indicated by A and B. Point A: Mono-static with AP1 to bi-static with APs 1 and 3. Point B: Bi-static with APs 1 and 3 to bi-static with APs 1 and 2.}}
    \label{fig:interference scenario}
\end{figure}

We consider two different scenarios for the simulations. First, in Fig. \ref{fig:sensing scenario}, we consider an interference-free sensing scenario, without communicating \acp{UE}, and with blockages that limit the sensing coverage area of the APs as illustrated in Fig. \ref{fig:system model}. However, in Fig. \ref{fig:interference scenario} and Table \ref{tab:simulation 3}, two UEs are added to the scenario, but the blockages are not considered and we analyze the effect of the sensing handover on the communication and sensing performance.
\begin{table*}[t]
    \centering
    \caption{Simulation parameters}
    \begin{tabular}{|c|c|c|c|}
        \hline \textbf{Number of APs} & $3$& \textbf{Channel model} & Geometric channel model\\\hline
        \textbf{Number of UEs} & $2$ &  \textbf{non-LOS model}&Rician model \\\hline
        \textbf{AP antennas}& $64$ & \textbf{$K$-Rician} & $-5$dB\\\hline
        \textbf{UE antennas} & $1$ & \textbf{Beamforming technique} & DoD/DoA beamforming\\\hline
        \textbf{Frequency} & $24$ GHz  & \textbf{Path-loss model} & Log-N model\\\hline
        \textbf{AP transmit power} & $40$ dBm &\upd{\textbf{Path-loss exponential LOS and NLOS}} & \upd{$2.1$ and $3.1$}\\\hline
        \upd{\textbf{Noise power}} & \upd{$-60$ dBm}  &\textbf{Reference distance} & $1$m \\\hline
        \upd{\textbf{\ac{SI} power}} & \upd{$-45$ dBm}  &\upd{\textbf{Reference distance attenuation}} & \upd{$21$ dB} \\\hline
        \textbf{Snapshots} & $120$ &\textbf{Radar RCS model} & Swerling 1 model\\\hline
    \end{tabular}
    \label{tab:parameters}
\end{table*}

Figure \ref{fig:sensing scenario} presents the received \ac{SNR} of the object echo for all possible combinations of the mono-static and bi-static sensing as well as the achievable sensing SNR in the cases with sensing handover enabled. Here, the hard sensing handover process is considered, where a sensing handover is triggered as soon as a new mono-static or bi-static sensing configuration with better received sensing SNR is detected. Moreover, the Tx/Rx functionality of the APs is adapted to maximize the sensing SNR, i.e., at each point, the optimal Tx/Rx configurations of the nodes are considered such that the sensing SNR is maximized. Due to the full-duplex challenges, such as strong \ac{SI}, the AP antennas are divided in half, allowing for simultaneous pulse transmission by the first half and reception of the object's echos by the other half. \upd{In practice, simply dividing the antennas is not enough to mitigate the SI, for this reason, if an AP is operating in full-duplex mode we also consider an interference of $-45$ dBm to represent the SI.}

As illustrated in Fig. \ref{fig:sensing scenario}, initially, only AP1 has LOS to the object, and the network resorts to mono-static sensing with AP1 to locate the object.  
At snapshot $25$, the object enters the AP$2$'s coverage area, allowing for bi-static sensing between APs $1$ and $2$, triggering the handover process with an increased SNR of $5$ dB. Here, the network starts the handover process from mono-static to bi-static sensing (Point A in Fig. \ref{fig:sensing scenario}).  
With proper handshaking between AP1 and AP2, the previous estimations of the object's direction, measured by AP1, may be exploited to facilitate the object's localization at AP2. 

At snapshot $45$ (Point B in Fig. \ref{fig:sensing scenario}), the AP1-AP2 bi-static sensing SNR decreases because the object has moved outside the AP1's sensing coverage area, triggering the handover process from bi-static to mono-static sensing using AP2 which has shown to have the best performance among others sensing configurations. In this case, AP1 stops the sensing process, and AP2 starts transmitting sensing pilots with half of its antennas while the other half receives the object's echoes. In this case, the transmission and reception beams can be easily selected since AP2 has already sensed the object and has \textit{a-priori} information about the object's moving trajectory. Finally, a similar process happens at snapshots $72$ and $84$ (Points C and D in Fig. \ref{fig:sensing scenario}); first, we have the sensing handover from AP2 mono-static to bi-static with AP2 and AP3, increasing the SNR by $5$ dB, and secondly, the sensing is handovered from bi-static with APs 2 and 3 to mono-static with AP3 avoiding a $50$ dB SNR drop. In this way, sensing handover guarantees a sensing SNR above $10$ dB during the moving trajectory, while without sensing handover, none of the considered sensing configurations can provide uninterrupted sensing of the object.
 
In Fig. \ref{fig:interference scenario}, we consider that the object's trajectory is within the coverage area of all APs, and study both the communication and sensing performance in the presence of UE1 and UE2, as illustrated in Fig. \ref{fig:system model}. The UEs are assumed to be stationary while the object follows the moving trajectory given in Fig. \ref{fig:system model}. 
\upd{AP2 and AP3 communicate with UE1 and UE2, respectively, in an DL slot. It is assumed that the communications links do not cause interference with each other since, in practice, traditional multiple-access techniques can be used to mitigate this interference. Nonetheless, the interference between the communication and sensing links is considered.}
Here, we consider a setup where a minimum quality-of-service (QoS), represented by the communication \ac{SINR} of $15$ dB, \upd{needs to be satisfied for both UEs.} Otherwise, the sensing is handovered to other nodes. Such a setup is of interest because, in practice, communication may have higher priority over sensing in an ISAC network.

Initially, AP1 operates in mono-static sensing mode to track the object, while APs 2 and 3 operate in communication mode. With that strategy, \upd{the communication links have high SINRs since the object and the UEs are far from each other. On the other hand, the sensing performance is poor because of the \ac{SI}, present in full-duplex operation mode, as well as the interference caused by the communication links.}
However, at snapshot $25$ (Point A in Fig. \ref{fig:interference scenario}), the object and UE1 are aligned with respect to AP2, causing considerable S2C interference. Thus, UE1's SINR drops, which triggers the sensing handover. Here, the network starts the sensing handover from \upd{mono-static with AP1 to bi-static with APs 1 and 3. However, the UE1 SINR is still below the threshold for a few snapshots where the object and UE1 are aligned. In fact, since AP1 is now using all the antennas to transmit, the S2C interference increases, reducing the communication SINRs in both UEs, but due to the distance between AP1 and UE2, the communication SINR drop is less pronounced. Regarding the sensing SINR, a slight improvement is noticed since the transmitter uses more antennas to transmit, but the sensing receiver is operating in full-duplex and also suffering from SI.}
\upd{At snapshot 104 (Point B in Fig. \ref{fig:interference scenario}), as the object moves closer to AP3, the S2C interference increases, and the communication SINR at UE2 drops below the threshold, which triggers the sensing handover from bi-static with APs 1 and 3 to bi-static with APs 1 and 2. After the handover, UE2 SINR increases slightly since AP3 is only operating in communication mode with UE2, being able to use all antennas to transmit communication data. On the other hand, UE1's SINR reduces significantly because AP2 is operating in full-duplex, transmitting communication data to UE1 and receiving reflections of sensing pilots from AP1, but UE1 is still operating above the desired level. The sensing performance is enhanced considerably, since the C2S interference from AP3 to AP2 is weaker than the C2S interference from AP2 to AP3 in the bi-static AP1 to AP3.}
In this way, sensing handover enables proper sensing of the object during the moving trajectory while the UEs' minimum QoS requirements are satisfied \upd{for most of the observations.}    

In Table \ref{tab:simulation 3}, we study the average sensing performance for the object moving in the trajectory of Fig. 2 in the cases with and without sensing handover and random UE positions. Particularly, the table demonstrates the probability that the sensing SINR is above a threshold \upd{$X=5, 10, 15$} dB as the object moves in the area. Also, the table shows the expected number of required handovers in the cases where sensing handover is possible.
In this scenario, two UEs are randomly dropped at each iteration, and each UE is connected to one AP, such that there are always two APs serving the UEs, and one AP is available to sense the object. In this case, the handover disabled method in Table \ref{tab:simulation 3} resorts to mono-static sensing with the available AP constantly. Also, success probability indicates the probability that the sensing SINR remains above the considered threshold.

As shown in Table \ref{tab:simulation 3}, the sensing handover guarantees that, with high probability, the sensing SINR remains above the desired threshold, while without sensing handover, the minimum required sensing SINR may not be satisfied for a large portion of the moving trajectory. Moreover, the expected number of required sensing handovers increases in the cases with strict sensing SINR requirements. However, depending on the UEs' positions and the sensing QoS requirements, few sensing handovers may be required during the object's movement.


\begin{table}[t]
    \centering
    \caption{Average performance comparison between the sensing handover strategy and baseline, without handover, for random UEs positions and different minimum sensing SINR threshold values $X$.}
    \resizebox{\columnwidth}{!}{%
    \begin{tabular}{|c|c|c|c|}
    \hline
        \textbf{\makecell{Sensing SINR \\ threshold $X$}}& \textbf{\makecell{Success probability \\ handover enabled}} & \textbf{\makecell{Average number \\ of handovers}} & \textbf{\makecell{Success probability \\ handover disabled}}\\\hline
        $5$dB & $99.9$\% & $0.002$ & $99.8$\% \\\hline
        $10$dB& $99.9$\% & $0.004$ & $91.3$\% \\\hline
        $15$dB& $99.2$\% & $1.002$ & $60.4$\%\\\hline
    \end{tabular}}
    \label{tab:simulation 3}
\end{table}

\section{Conclusions}
This paper studied the potentials and challenges of mobility management and sensing handover in ISAC networks. As demonstrated, with proper configurations, sensing handover guarantees uninterrupted sensing, reduces signaling overhead/network interference, and improves the network knowledge about the area, which, in turn, improves the communication performance. 

It is interesting to note that the standardization process regarding ISAC has started recently in 3GPP in a study item.
 As a result, there is no standardization related to sensing handover or mobility management in ISAC networks. Moreover, while we concentrated on connected mode mobility, using idle mode signals for sensing mobility management may be even more challenging. Hence, even if ISAC or mobility management of ISAC are agreed to be included in standardization, extensive standardization work is required before
ISAC can be used in practical wireless networks. 

\bibliographystyle{IEEEtran} 
\bibliography{Ref}


 
%


\section{Biography Section}
 



\begin{IEEEbiographynophoto}{Yuri S. Ribeiro} (Student Member, IEEE)
is with the Wireless Telecommunications Research Group (GTEL), Universidade Federal do Ceará (UFC).
\end{IEEEbiographynophoto}
\begin{IEEEbiographynophoto}{Behrooz Makki}
(Senior Member, IEEE) is with Ericsson Research, Ericsson,
417 56 Goteborg, Sweden. 
\end{IEEEbiographynophoto}

\begin{IEEEbiographynophoto}{André L. F. de Almeida}
(Senior Member, IEEE) is with the Wireless Telecommunications Research Group (GTEL), Universidade Federal do Ceará (UFC).
\end{IEEEbiographynophoto}
\begin{IEEEbiographynophoto}{Fazal-E-Asim}
(Senior Member, IEEE) is with the Wireless Telecommunications Research Group (GTEL), Universidade Federal do Ceará (UFC).
\end{IEEEbiographynophoto}
\begin{IEEEbiographynophoto}{Gabor Fodor}
(Senior Member, IEEE) is with Ericsson Research, Ericsson,
16480 Stockholm, Sweden, and also with the School of Electrical Engineering
and Computer Science, KTH Royal Institute of Technology
\end{IEEEbiographynophoto}


\end{document}